\documentclass[fleqn,twoside]{article}
\usepackage[headings]{espcrc2}

\readRCS
$Id: espcrc2.tex,v 1.2 2004/02/24 11:22:11 spepping Exp $
\usepackage{graphicx, balance}
\usepackage[figuresright]{rotating}
\usepackage{hyperref}
\usepackage{multirow}
\usepackage{enumerate}
\usepackage{amsmath}
\usepackage{tikz}
\usetikzlibrary{trees,shapes}
\usetikzlibrary{decorations.pathmorphing}
\usetikzlibrary{decorations.markings}
\usetikzlibrary{arrows}

\title{\textbf{Object Oriented Analysis using Natural Language Processing concepts: A Review }}

\author{Abinash Tripathy\address[DCSE]{Department of Computer Science and Engineering, National Institute of Technology, Rourkela, Odisha, India, Contact: abi.tripathy@gmail.com \\},
{Santanu Kumar Rath\address[DCSE]{Department of Computer Science and Engineering, National Institute of Technology, Rourkela, Odisha, India, Contact: skrath@nitrkl.ac.in \\}}}

\runtitle{Object Oriented Analysis using Natural Language Processing concepts: A Review}
\runauthor{Abinash Tripathy, et al.}

\begin{document}
\begin{abstract}
The Software Development Life Cycle (SDLC) starts with eliciting requirements of the customers in the form of Software Requirement Specification (SRS). SRS document needed for software development is mostly written in Natural Language(NL) convenient for the client. From the SRS document only, the class name, its attributes and the functions incorporated in the body of the class are traced based on pre-knowledge of analyst.
 \par
The paper intends to present a review on  Object Oriented (OO) analysis using Natural Language Processing (NLP) techniques. This analysis can be manual where domain expert helps to generate the required diagram or automated system, where the system generates the required digram, from the input in the form of SRS.

\end{abstract}

\maketitle
\section{Introduction}
\label{section-1}
Software Requirement Specification (SRS) document forms the basis of problem analysis between client and developer. SRS needs to be very specific, while serving as a basis, to proceed towards implementation of desired software. It is very often observed that SRS is expressed in any natural language as comprehensible by the client. But it may be ambiguous, possibly inconsistent, and probably unmanageably large from the software analyst's point of view.
\par
Identifying major functionalities from the OO analysis point of view plays an important role in project success. The use of formal languages like Unified Modeling Language (UML) have been applied to avoid the inherent problems of natural language such as incompleteness and ambiguity \cite{rumbaugh}.
\par
Earlier analysis was used to help for an explanatory model called as build and fix programming style. But this style was observed to be very informal and there are no set of rules as to which one is superior. Every programmer formulates his own software development technique solely guided by his expertize and in his own language and style \cite{pressman}.
 \par
   In recent years, the object-oriented software development style is a preferred style over conventional style by developers as the present day software development languages are object oriented in nature. Hence, OO analysis of software helps to find out the candidate for class, function, and the attributes associated with those classes. 
 \par 
  Natural Language Processing (NLP) combines the effect of computer science and linguistics branch which are concerned with the interaction between the computer and human languages \cite{kumar}. Natural Language generation systems mostly extracts right information from statements which are in human readable form.
  \par
The aim of the work is to present a review on existing literature of application of NLP in Object Oriented Analysis (OOA) based on literature available such as: Abbott \cite{Abbott},   Saekai and Enamoto \cite{Saeki}, Nanduri and Rugaber \cite{nanduri}, Juristo and Moreno \cite{juristo}, D. Popescu et al. \cite{popescu2008reducing},  Ibrahim and Ahmed \cite{ibrahim},Harmain and Gaizauskas \cite{harmain2000cm}, Overmyer and Rambow \cite{overmyer2001conceptual}, Mich \cite{mich2002nl}. Among these literatures few authors suggested automated tools, other use manual approach and few combine both tool and manual approach to obtain the different elements of OO analysis .
\par

In order to examine each proposal, the following dimensions may be considered

\begin{itemize}
\item \textbf{Steps to modify the SRS into required form:} As discussed above the SRS is written in a form i.e., convenient to the user. During this step the SRS is modified to a format by which the process of finding the keywords became easier for analysts.

\item \textbf{Finding out the candidate for class and object from modified SRS:} After transforming the SRS to the required format for analysis, then the candidates for the class name and its details are traced out. The process of identification of the class name and its detail can be a manual or a automated process. In manual process, the domain expert analyzes text to bring out ``intermediate output'' then automated process considers the intermediate output to generate the desired output.
\end{itemize} 

The objective of this work is two-fold:
\begin{enumerate}
\item This analysis provides information on available techniques for use of NLP, further to be considered under OO analysis.
\item It provides an overview of the current state for the use of NLP in OO analysis, focusing on the strengths and weaknesses of existing proposal. Thus researchers can have a broad knowledge into the work that already being done and that still can be carried out  in this field.
 \end{enumerate}

The paper is organized as follows: In section 2 the existing proposals of the use of NLP for OOA analysis are presented. Section 3 presents an analysis on all proposals. Finally, section 4 discusses on some concluding remarks and highlights the future trend.
   
\section{Proposal of the use of NLP in OO Analysis }
\label{sec-2}
\subsection{Historical review on NLP }
 Jones in his paper presented a review paper on NLP based on the historical study \cite{jones}. The paper reviews study of NLP from late 1940's to present by distinguishing four phases in the history of NLP.  The impact of use of machine translation, artificial intelligence influence, logico-grammatical style adaption and massive language data attack act as the basis of the phase division.  
 
\begin{enumerate}

\item [\textbf{Phase 1.}]
The early phase of study on NLP was during \textbf{late 1940s to late 1960s} and in this period focus was mainly in Machine Translation (MT). Noticeable amount of work done in USSR, USA, Europe and Japan during this period. Thus the languages considered for research in this period were mostly Russian and English \cite{booth}. Language syntax was mainly the area of research  in this period as syntactic processing was manifestly necessary, and partly carried out through implicit or explicit endorsement of the idea of syntax-driven processing. Though during this period use of computers for literary and linguistic study has began, but it has never been linked with NLP.

\item [\textbf{Phase 2.}]
 Next phase of study was undertaken from \textbf{late 1960s to late 1970s} and the work mainly focused on use of artificial intelligence (AI) in NLP, with much more priority on word knowledge and on its implementation in the construction and manipulation of meaning representations. AI was mainly considered during this period for construction and addressing of knowledge base or data. In late 1960s, the prevalent theory of linguistic is transformational grammar, which provides the semantic information about NLP. 
 
\item [\textbf{Phase 3.}]
This phase was mainly concerned to the period of \textbf{late 1970s to late 1980s} and characterized as grammatico-logical phase. The requirement of development of grammatical theory and movement towards incorporation of logic in knowledge representation and reasoning  triggered during 70s. During this phase, deliberate attempts made to transform the commercially available dictionary  to  machine-readable form which further helps in text corpora validation and customizing lexical data.

\item [\textbf{Phase 4.}]
 The forth and the final phase can be attributed as the study carried out from \textbf{late 1980's onwards}. During this phase, the main area of research is statistical language data processing. The identification of linguistic occurrence and patterns in the corpus for both  syntactic and semantic analysis, drew interest in this period. The present attention on lexicon, retrieving statistical information, and restore interest in MT. 
 \end{enumerate}
 
Table \ref{time} compares the NLP research based on time period

\begin{table}[h!]
\caption{Comparison of NLP research based on time }
\label{time}
 \renewcommand{\arraystretch}{1.5}
             \resizebox{7cm}{!}{
\begin{tabular}{|c|p{2cm}|p{5cm}|}
\hline
Phase   & Time period              & NLP research \\ \hline
Phase 1 & Late 1940s to late 1960s &  Machine Translation (MT), Language syntax analysis.            \\ \hline
Phase 2 & Late 1960s to late 1970s & Use of AI, implement of word knowledge to construction knowledge base or data.            \\ \hline
Phase 3 & Late 1970s to late 1980s  & Grammatico-logical analysis, Transformation of dictionary to machine readable form and text corpora validation.            \\ \hline
Phase 4 & Late 1980s onwards &   Statical data processing, Both syntactic and semantic analysis of Corpus and restoration of MT.            \\ \hline
\end{tabular}}
\end{table}

\subsection{OO Analysis using NLP approaches}
In course of this section, the study made by various authors are analyzed on the basic of how they transform the SRS and how the candidate for the class name and its details are found out from transformed SRS.

\subsubsection{\textbf{Abbott, 1983} \cite{Abbott}} 
\label{sec-2.1}
The study made by \textbf{Abbott} proposes a method to derive the elements of object oriented analysis i.e., data type, variables, operators, attributes and candidate for class name form English statements. This paper shows the process of analysis of the English statement of SRS and helps to generate elements of OO analysis.

The approach used in this paper is divided into three different sections. These are as follows
 \begin{enumerate}[i.]
 
\item \textbf{Development of informal strategy for the problem:} The informal strategy should suggest the problem solution on the conceptual level. This step express the solution of the problem in terms of problem domain.

\item \textbf{Formalize the Strategy:} The  second  step  is
formalizing  the  solution  by finding out its  data  types,  objects,  operators,  and  control constructs. The steps of formalization are:
\begin{enumerate}
\item \textit{Identify the Data types:} The data types are suggested by the common nouns. The name of a class of beings or things are known as common noun. 
\item \textit{Identify the objects of those types:} Objects are suggested by Proper Nouns or Direct References. The name of specific things or beings is known as proper noun. A specific, previously identified being or thing without necessarily referring to it by name is known as direct reference.
\item \textit{Identify the operators for the objects:} Operators are suggested by verb, attribute, predicate or descriptive expression. Attribute is a property, association, characteristics or component of something. Predicate designates a property or relation that can be consider True or false i.e., to hold or not to hold. A descriptive expression is a characterization for which there may be some particular object.
\item The \textit{control structures} are directly provided by the English language.
\end{enumerate}
\item\textbf{ Segregate  the  solution  into  two  parts, A  package  and subprogram:} The package will contain the formalization of the problem domain, that is, the data types and their operators. Then subprogram(s) will contain the  specific steps (expressed in terms of the data types and operators defined in the package) for solving the particular  problem. 
 \end{enumerate}
 
During the course of the paper, different types of nouns are analyzed i.e., the difference of common noun with proper noun, direct references and mass nouns are provided. Classes of objects are referred by Common Noun but specific and individual objects are referred by Proper noun. Mass nouns are names of qualities, substances, and activity that do not have an a priori organization into individual units or instances. 

In this paper Abbott had taken an example of \textit{``Calculating the days between two given dates''} to explain his technique. As per the analysis referred in the paper, the process is divided into three different steps:
\begin{itemize}

\item In step 1, an informal strategy of the problem analysis is provided. In this step the process of getting the detailed solution is being analyzed. 

\item In step 2, the data types, objects, operators, and control structures are found for the specific problem form the informal specification.

\item In step 3, the final solution is being prepared. The package for the problem is assigned and the subprogram details are provided in this step.
\end{itemize}
Though this paper is comparatively easy in solving the problem of finding the candidate for class name and its details but the process is manual. A software engineer having a good knowledge about the domain requires to provide the step-wise solution of the problem. 

   
\subsubsection{\textbf{Saeki et al., 1989} \cite{Saeki}} 
The paper by \textbf{Saeki et al.} discusses the process of derivation of formal specification from an informal specification written in natural language. The informal specification contains important information leading to their formal specification or the prototype program. Then the similarity between the structure of the words and the structure of software component is analyzed. In this paper, the \textbf{``Lift Control System''} example has been explained as an informal specification to explain the process. The process consist of three major steps as follows:

\begin{enumerate}[i.]
\item \textbf{Design Activity:} The purpose of this step is to construct a module design document from informal specification. The modular design document presents the modular structure of a formal specification, i.e., external design of class modules which contain class names, method names and message protocols. This design activity consists of several sub-activities. Each of them produce an intermediate product using the informal specification. The intermediate products obtained are as follows:
\begin{itemize}
\item \textbf{Noun table}: This table contains information about extracted nouns. According to the author, noun can be classified into different group, i.e., \textit{Class noun} identifies object or set of objects, \textit{Value noun} identifies the values or set of values, \textit{Attribute noun} identifies the attribute of the objects and \textit{Action noun} identifies the actions to be carried out.

\item \textbf{Verb table:} This table contains information about extracted verbs, i.e., verb names, their categories, their subjects and objectives. According to the author, verbs can also be classified into different groups. \textit{Relation verb} specifies the relation between objects or between objects and their attributes. \textit{State verb} specifies the internal state of the object or the attribute values of the object. \textit{Action verb} specifies the action to the objects and \textit{Action relational verb} specifies the relation between the actions.

\item \textbf{Action table}: This table presents the extracted actions, their agents, target objects and input output parameter associated with them. For each action verb, there is always an agent and its target object. The action verb changes the state of the target object. In order to extract a target object from a sentence, verb patterns that appear in various kinds of natural language specification are examined. 

\item \textbf{Action Relation table:} The informal specification, its verb table and its action verb are needed to identify the relationship between the actions. For every action to be performed, the sender, receiver and the message transmission between them need to be identified. in this paper, the authors have used a rule called ``action relation rule'' to generate a few candidate for a sender-receiver pair. 
 
\item \textbf{Module Design Document:} A module design document is constructed using above mentioned tables. The noun table helps to identify objects and their attributes. The verb table is used to extract relationship among objects and kind of attributes, each object possess. The module design document can have both graphical and textual representation. Both of them are based on syntax of formal specification language TELL and object oriented language Smalltalk80. 
 \end{itemize}

\item \textbf{Elaborate - Design Activity Cycle:} The task of this activity is to refine and rewrite the informal specification  as per module design document. The output of this activity is a natural language description which is accurate, detailed, structured yet informal. Whenever the elaborate-design cycle is carried out, a pair of informal specification and its module design document is obtained. The elaborate activity consists of sub-activities such as paste ,refine and an intermediate product a paste document is generated.
\begin{itemize}
\item \textit{Paste:} During this phase, the sentences of informal specification may be paraphrased to accurate sentence and then replace the original one. In the updated expression, the nouns and verbs are extracted as classes, and attributes or method to be used respectively.
\item\textit{ Refine:} The informal specification of each pasted module are constructed during refine activity. During this phase, the internal behavior and property of each class module is rewritten again, which are used for construction of each class. 
\item \textit{Design activity for elaborate Informal Specification:} Before this activity, the module design document for each class is constructed from elaborate informal specification. During this activity, each class module and method module are composed into small sub module to realize the internal behavior and property.
\end{itemize}
The cycle continues until a formal specification is obtained from the informal specification. The requirements need to be refined and made simpler and smaller during this cycle.

\item \textbf{Software process based on Natural Language:} During this process the design and elaborate process are embedded. Before this phase, the informal specification is already converted to formal specification, the rest steps are as follows
\begin{itemize}
\item \textit{Analyze activity: }This step acquires a problem description by means of  interaction between customer and developer. An informal specification is obtained as an output of this step.
\item \textit{Evaluate activity:} During this activity, the obtained formal specification is executed and verified. The diagnostic document keeps a record of the execution and verification.
\item \textit{Evolve activity:} During this activity, the developers check the diagnostics document and find whether something is to be modified or not. If so, they create a new version of informal specification, this process is called as ``evolve activity''. From this new document, new module design document, new formal specification, new diagnosis document are produced. 

\item \textit{Instantiate activity:} When the developers judge that the formal specification is accepted to the customer's need, they finally produce the concrete program code. This activity is called as instantiate activity. A formal specification is considered to be a generalization or abstraction of program code. 
\end{itemize}
\end{enumerate}

In this paper, the author has presented a software design process based on natural language and obtained formal specification from informal one through the design and elaborate activity. The technique used in this paper is verb-oriented which has an impact on the dynamic nature of informal specification. Along with this, nouns also needed to extract the hierarchy of class.

\subsubsection{\textbf{D. Popescu et. al.,2008} \cite{popescu2008reducing}}
This paper by \textbf{D. Popescu et. al.} proposed an approach to help the writer or reviewer in identifying the ambiguities in  NL SRS. A tool named ``Dowser'' is proposed by them which creates OO digram from NL SRS. Their approach consists of three steps
\begin{enumerate}[Step 1:]
\item The NL SRS is parsed according to a constraining grammar. 
\item From the obtained relationship after parsing,
the tool creates the elements of object-oriented
analysis model of the specified system.
\item The diagram of the model is generated, which is reviewed by human reviewer to detect any inconsistency or ambiguity.
\end{enumerate}

The authors in this paper prefer the use of model over analysis of the whole NL SRS due to following reason
\begin{itemize}
\item  All software companies irrespective of their domain use NL SRS for description of software system.
\item An Object Oriented Analysis Module (OOAM) shows the concepts and the relationship among them.
\item The OOAM model for each sentence is identified while OO design selects parts of the sentences, such as class from subject, attributes from adjective and methods from verb.
\end{itemize}

During the course of the paper, the authors had used few concepts such as
\begin{itemize}
\item \textbf{Constraining Grammar:} It is different from the internal grammar that parser uses. The constraining grammar attempts to reduce the number of way a statement can be represented and also try to make it uniguous (not ambiguous). On the other hand, the NL parser grammar checks the legitimacy of the sentence only, without checking whether it is ambiguous or not. 
\par This paper used the constraining grammar proposed by   
Juristo et.al., in their paper \cite{juristo}. It attempts to generate an unambiguous mapping form this grammar to OOAM. The constraining grammar influences the structure of NL SRS as it uses simple sentence consisting of subject, object and verbs.

\item \textbf{Natural Language Parsing :} As OOAM is generated using syntactic information automatically, parsing of NL SRS is needed to obtain the required information. The parser uses the link-grammar that consists of set of words. These words act as  terminal symbols and have different liner requirements.
\item \textbf{Transformation Rules:} These rule helps to transform the obtained syntactic information to targeted OOAM. Dowser have thirteen different transformation rules, the most generally used rules are:
\begin{itemize}
\item If the sentence contains both subject and object link after parsing, then two different classes are created with association named as verb.
\item To find aggregation, if the parsed sentence consists of both subject and object link and the verb is one of  ``have, posses, contain or include'' then the object is aggregated to subject.
\item ``if or when '' always represent the start of an event. If an event is detected by Dowser and the main clause only consists of subject link, then class is created with the noun present in subject link and verb acts as a method to it.
\item If genitivity detected by Dowser, two different classes are created with one linking  aggregation. In order to fix whether both became aggregated class or one became attribute of other, both syntactic and semantic information of the sentence are needed.
\item Though active clauses preferred in NL SRS still passive clause helps to describe relations and states. The passive verb and its connecting word describe the association.
\end{itemize}
Dowser applied two post processing rules after transforming
NL SRS. These are
\begin{enumerate}[i.]
\item It converts all classes that aggregated to another class as attribute of other class.
\item It removes class ``system'' from OOAM.
\end{enumerate}
\item \textbf{Domain-Specific Terms(DST):} Special domain data dictionary is needed by the parser to interpret the DST. But for each domain such dictionary does not exist. Hence, to improve the DST recall, the link grammar has a guessing mode that uses the syntactic role of unknown terms.
\item\textbf{Diagramming OOAM:} The textual OOAM is  created using previous steps. The tool \textit{UMLGraph} transforms the textual information into \textit{a dot file} while then transforms into graphic format using \textit{Graphviz} tool.
\item\textbf{Interpretation of OOAM:} In this step, a human analyst checks for ambiguity in generated diagram. The defects that can be found out in OOAM are as follows
\begin{itemize}
\item An association can be ambiguous; so, the analyst checks whether different classes transmit message to same class or not.
\item The classes should represent only one concept. As Dowser does not allow generalization principle, the concepts such as \textit{cash payment} and \textit{on-line payment} are not combined to form payment as a whole; but can be identified as two different classes.
\item If the classes have attributes that are not of primitive type, proper definition added to it so, it can be represented in own class.
\item The class must be associated with other class otherwise it becomes unspecified.
\end{itemize}
\end{itemize}

This paper only supports the static behavior/ relationship of OOAM present in NL SRS. It does not manage the modeling behavior.

\subsubsection{\textbf{Ibrahim and Ahmad, 2010} \cite{ibrahim}} 
This paper of \textbf{Ibrahim and Ahmad} proposes method to facilitate requirement analysis process and extraction of class diagram from requirements using NLP and Domain Ontology. A tool named `` Requirements Analysis and Class Diagram Extraction (RACE) '' is being proposed by the authors that analyzes the textual requirements, finding out the relationships and finally extracts the class diagram.
\par The RACE system consists of different internal and external components or sub-systems. These can be described as follows:

\begin{enumerate}[i.]
\item OpenNLP Parser: The OpenNLP parser used in this paper for lexical and syntactic parsing. The parser takes English text as input and  provides corresponding POS tag for each word as output.
\item RACE Stemming Algorithm: Stemming is a process of removing affixes and suffixes from a word and generating the base word. The generated base word reduces the redundancy and increases efficiency.
\item WordNet: It is used to validate the semantic of the sentences that generated after syntactic analysis. It also helps to display hyponyms for a selected noun, which helps to know the ``a kind of'' relationship. 
\item Concept Extraction Engine: This module is used to extract concepts according to the requirement document. The algorithm for this module is as follows:
\begin{enumerate}[Step 1.]
\item Requirement document is taken as input.
\item Stop words are identified and stored as \textbf{Stop-words\_Found} list
\item Calculate the frequency of each words in the document except the Stop words.
\item Use RACE stemming algorithm to stem each words and store them in a list.
\item Use OpenNLP to parse whole document.
\item From the parsed output extract the words with POS Proper Nouns (NN), Noun Phrases (NP), verb (VB) and store them in  Concept-list.
\item For each concept in concept-list, if any other concept is synonym with present one, then it can be conveyed that both are semantically related.
\item For each concept in concept-list, if any other concept is item Requirement document is taken as input.
\item Stop words are identified and stored as \textbf{Stop-words\_Found} list
\item Calculate the frequency of each words in the document, except the Stop words.
\item Use RACE stemming algorithm to stem each word and store them in a list.
\item Use OpenNLP to parse whole document.
\item From the parsed output extract the words with POS Proper Nouns(NN), Noun Phrases (NP), verb (VB) and store them in  Concept-list.
\item For each concept in concept-list, if any other concept is synonym with present one then it can be conveyed that both are semantically related.
\item For each concept in concept-list, if any other concept is hyponyms with present one i.e., lexically same then it can be conveyed that former is a kind of later and saved in  Generalization-list. with present one i.e., lexically same then it can be conveyed that former is a kind of later and saved in  Generalization-list.
\end{enumerate}
\item Domain Ontology: It is used to improve the performance of concept identification. In RACE system Library system ontology is being used. XML is being used to build the ontology.
\item Class Extraction Engine: The input to this module is the output of ``Concept Extraction Engine''. During this step, some heuristic rules are used by the authors to extract the class diagram. The rules are as follows
\begin{itemize}
\item Class Identification Rules: The rules used for extraction of classes are:
\begin{itemize}
\item If the occurrence of the concept is only one or frequency is 2\%, then the concept is ignored as class.  
\item If the concept is related to design elements, location name or person name, then ignore as class.
\item If the concept found in high level of hypernyms tree or an attribute then ignore it as a class.
\item If the concept is a noun phase and the second part is an attribute then consider the first part for class name.
\end{itemize}
\item Attribute Identification Rules: The rules for attribute identification is as follows
\begin{itemize}
\item  If the concept is a noun phase including underscore between two nouns, then the first noun is a candidate for class name and the second part  is attribute of that class.
\item If the concept has only one value then it is an attribute.
\end{itemize}
\item Relationship Identification Rules: The rules for relationship identification is as follows:
\begin{itemize}
\item Using step 8 of concept extraction engine, the element of generalization-list transferred as Generalization (is-a) relationship.
\item If there exists a sentence having (CT1-VB-CT2) where CT1 and CT2 are classes, then VB is an association rule.
\item If the sentence is of the form CT1+R1+CT2+``AND''+CT3	where CT1, CT2 and CT3 are classes and R1 is the relationship, then there exists relationship between (CT1,CT2) and (CT1,CT3).
\end{itemize}
\end{itemize}
\item RACE Concept Management: User interaction is important in RACE system. The UI helps to perform tasks such as creating and printing requirement and acts as an interface to add, modify, view and organize relationships.
\end{enumerate}
The RACE system is implemented using C\#, MS Access is being used for database operation, and to open textual requirements  word document, text file, rich text file, and HTML file are being used.

\subsubsection{Overmyer and Rambow,2001 \cite{overmyer2001conceptual}}

This paper of \textbf{Overmyer and Rambow} proposes a tool called Linguistic assistant for
Domain Analysis (LIDA) that  provides linguistic assistance for model development process. This tool helps to obtain the OO model for a domain using UML. In order to perform this task, large volume of text from ``Legacy system'' is collected. The LIDA tool considers the following features
\begin{itemize}
\item Domain independent linguistic processing used to group different form of base words using POS and to find multi-word terms.
\item The final output is in the form of full text, word-list and UML model in parallel. So the user can compare all of them.
\item Key Word In Context `KWIC' view displays the words or group of words in sentences.
\item Hypertext description model used to help in documentation of  the model.
\item  Completed model can be exported any CASE tool or any model can be imported from any CASE tool to LIDA.
\end{itemize}

LIDA consists of following components:
\begin{enumerate}[i.]
\item \textit{Text analysis environment:} This component is the main component of LIDA as it provides the central functionality. The main functionalities this component performs are:
\begin{itemize}
\item It takes the text input in RTF and ASCII format.
\item Then it assigns POS tag to each word. For POS tagging, It uses MXPOST, a software tool developed at University of
Pennsylvania,USA.
\item Base word is obtained form each word and their frequency is calculated.
\item  Multi-word phases are checked for a given base word.
\item Users are allowed to mark the words or phases as candidate model and highlights these words in the text .
\item Retrieve textual context of marked words.     
\end{itemize} 

 \textit{Mode editing environment:} This model offers the functionality requirement to generate a model from the proposed model marked in LIDA’s Text Analyzing Environment. The functional features of this components are:

 \begin{itemize}
 \item \textbf{Display} list of candidate model element marked and add them to model editing environment. Transfer of information between text analysis environment and  mode editing environment helps developer to analyze the problem in details.
 \item \textbf{Suggest} operations for combining elements to class model.
 \item \textbf{Add} textual context helpful for processing model builder.
 \item \textbf{Generate} textual description of model for documentation and validation of model with domain expert.
 \end{itemize} 
 
 \item \textit{LIDA text description:} LIDA uses Model-Explainer an integrated tool, which generates the hypertext description document for object model. This document is generated from customized text which includes the class information like superclass, subclass, attributes, operation and association with other class. These descriptions help to obtain additional information about the final result. 
\end{enumerate}

The following Table \ref{review} provides a comparative analysis of the approaches to obtain the elements of OO analysis from SRS.
 
 \begin{table}[h!]
 
 \centering
 \caption{Comparison of generation of OO elements from SRS using manual approach}
 \renewcommand{\arraystretch}{1.2}
            \resizebox{8cm}{!}{
 \begin{tabular}{|l|p{5cm}|p{4cm}|p{4cm}|}
 
 \hline
 Authors   &  Proposed Approach & Advantage & Limitation  \\ 
 \hline
 Abbott \cite{Abbott}& This paper analyzes the English statement of SRS and generate elements of OO analysis. Identifying data type, objects, operators and Control structure & Comparatively easy to find out the candidate for class and it's details & Domain knowledge is required for the analysis.   \\ 
 \hline
Saeki \textit{et al.} \cite{Saeki} & This paper derive formal specification from informal specification in English and from that obtain elements of OO. Generate Noun table, Verb table, Action table, Action Relation table  and Module Design document from formal specification & The informal SRS document is refined and rewritten to a formal document understandable for everyone. & The large size of informal specification may be a concern and also further analysis on nouns needed as the  proposed approach is mainly verb oriented.      \\
 \hline

Nanduri and Rugaber \cite{nanduri}& This paper extract the candidate objects, methods and its association from requirement document then composing them to generate object model. Use  link grammar based parser to parse sentence and generate the object diagram from knowledge gained.  & A graphical model is generated from the requirement document & Parser inadequacy, ambiguous and incomplete specification and lack of domain knowledge makes the final result unsatisfactory. \\
 \hline
 Juristo et. al.\cite{juristo}&  This paper uses the linguistic information from informal specification. Analyzes the information semantically and syntactically and finally apply semi-formal procedure to obtain OO system component. & The proposed approach prevent incorrect modeling construct and model can be repeatable  & As the process totally depend on requirement specification, an assumption taken that the textual document is correct.   \\
 \hline
  D. Popescu et al. \cite{popescu2008reducing}& This paper identify the ambiguities in NL SRS. The proposed ``Dowser'' tool use constraining grammar, NL parsing and Transformation rule to generate the Object model. & The OOAM diagram is generated using tool, again verified by human analyst for better accuracy.  & Only the static behavior of SRS is considered, it does not manage modeling behavior.   \\
  \hline
   Ibrahim and Ahmad \cite{ibrahim}& This paper uses the requirement analysis process and extract class diagram using NLP and Domain Ontology. The proposed RACE tool analyzes textual requirements, finds relationship among them and finally generate class diagram. & RACE find the concepts based on nouns, noun phase and verb analysis. It can able to find generalization, association , composition, aggregation and dependency relationships. & It could not find out one to one, one to many or many to one relationship and RACE is not platform independent, it works only in windows platform. \\
    \hline
     Harmain and Gaizauskas \cite{harmain2000cm} & This paper uses CM-Builder tool for OO analysis. After analysis of software requirement document a discourse model is designed from which the object class and relationships is generated.  & The proposed model used different linguistic technique to analyze and define rule to generate candidate for class model. So, the ambiguities present in the software requirement document do not hamper the result.  &  The final output is obtained in CDIF form which is not understandable by everyone and a CASE tool supporting CDIF needs to generate class diagram graphically       \\
     \hline
      Overmyer and Rambow \cite{overmyer2001conceptual} & The paper uses LIDA tool to provide linguistic information assistance in model development process. This assistance facilitates the analysis and extent the creation of class model.  &  It provides a graphical approach to analyze the text and have features that can simplify the process of class generation.  & The text analysis carried out is mostly manual so it is time taking and the analyst should be a domain expert which is quite difficult to find out. \\
       \hline     
     
      Mich \cite{mich2002nl} & This paper uses an NL-OOPS tool based on LOLITA. The OO modeling module, use algorithms that filter the entity and event nodes generated by LOLITA and identify classes and associations. & It provides an graphical interface which make the process of generation quite easy. Again this tool can be very easily integrated to other CASE toolto support lower level development.  & In order to make the analysis fully automated, a senior analyst have to control the output and the final output class model is not at par with the  UML class diagram.    \\ 
      \hline

 \end{tabular}}
 
 \label{review}
 \end{table}
 
\section{ Analysis of Approaches} 

In present day scenario, the use of object oriented system is widely applied for software development. The customer mention all it's requirements in a document called Software requirement specification. This SRS document is written in NL which is understandable by the customer side, but it is sometimes incomplete and ambiguous.  The development team need to go through these document and generate UML diagram and analyze on basic of OO analysis. The UML being very often used for OOA tries to fix the class diagram where class is also basic element of OO system.

\par During the course of the paper, it can be found out that different approaches are adapted to generate the class diagram and its corresponding details. These approaches can be mentioned as follows:

\begin{itemize}
\item The software requirement document is considered as an input for the analysis.
\item As it is written in NL, it contains some ambiguities or unwanted information. So, in order to remove that different steps are carried out in all papers.
\item Each words in the text is tagged with a POS. Then the words are combined together depending upon their POS.
\item The noun and verb tagged words are mainly used for class name and their operations respectively. So these words are then stemmed to obtain the root word and their suffixes.
\item After obtaining the root noun words, the  higher frequency nouns are considered and they are the most eligible onces for fixing class name.
\item For operations in class, the verbs present in the sentence are the best candidate.
\item The Adjectives present in text act as an attribute for the class for that noun it tries to modify.
\item In order to find the relationship between classes, the relation between the subject and object of a sentence is found out.  
\item For other rules like multiplicity determines are used that specify the relationship like one-one, one-many, many-one, many-many.
\end{itemize}

 \section{Conclusion and Future scope}
 
 There are different tools that have been developed to analyze the text; but as there is no exhaustive dictionary which helps to provide POS for each words. Although few tools generate the class diagram but different authors suggest that a manual intervention is needed to improve the final result. Until and unless there is specific rules for writing  the SRS document, the ambiguities continue to be present in it and that cause issue in compiling the SRS.
 
\par  Though many approaches have been proposed and also are used to obtain the elements of OO analysis still there is scope for research in this area. To automated understanding the SRS written in informal NL is also an issue in research.

\noindent{\includegraphics[width=1in,height=1.7in,clip,keepaspectratio]{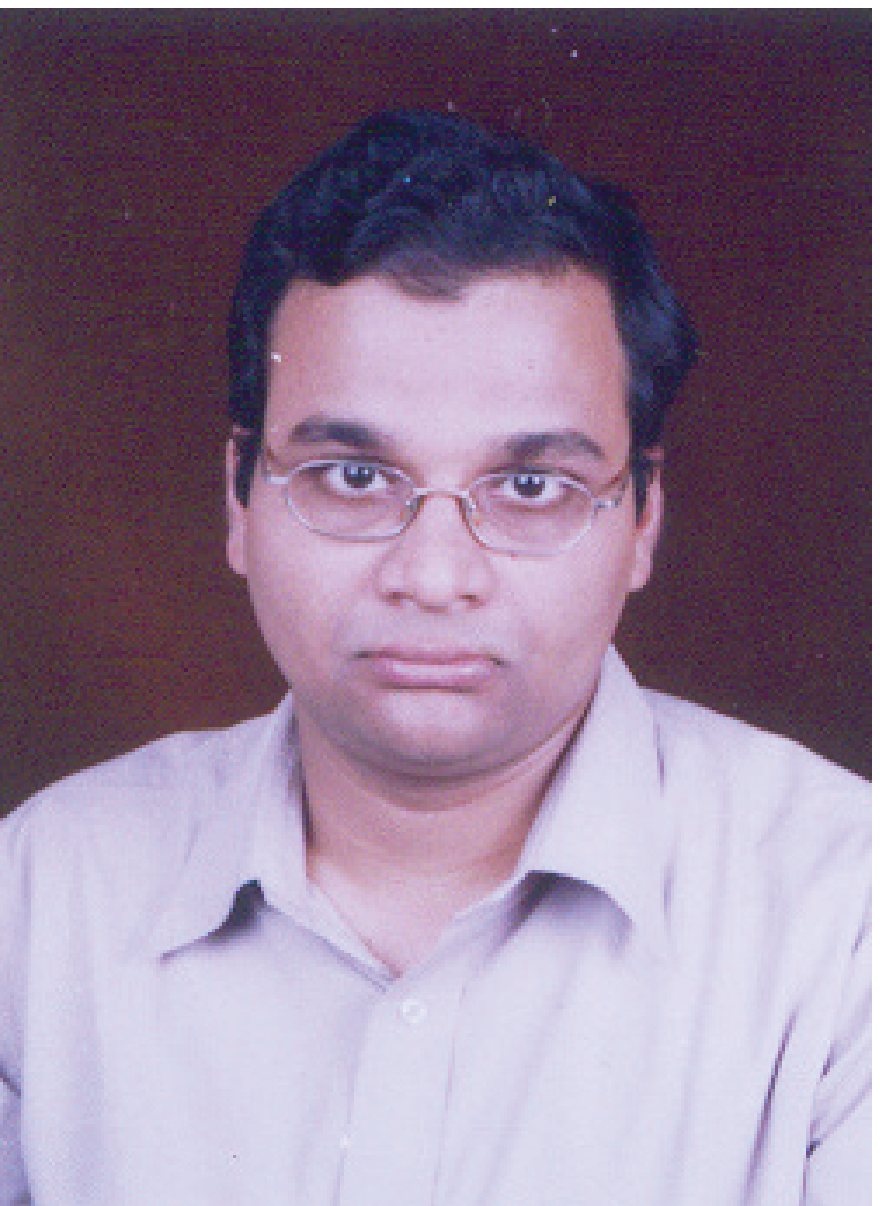}}
\begin{minipage}[b][1in][c]{1.8in}
{\centering{\bf {Abinash Tripathy}} is currently pursing his Ph.D. at National Institute of Technology, Rourkela. He obtained his Master degrees, M.Sc. Computer Science from Utkal University, Bhubaneswar and    }\\\\
\end{minipage}
M.Tech. Computer Science and Engg. from KIIT University, Bhubaneswar. His research interest are Software testing, UML, Natural Language Processing and Sentiment analysis.  \\\\
\noindent{\includegraphics[width=1.3in,height=2.3in,clip,keepaspectratio]{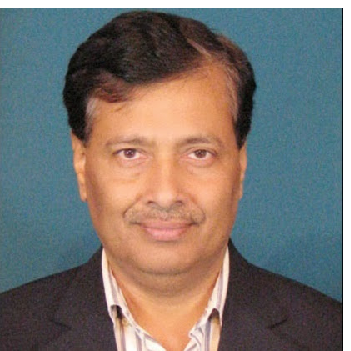}}
\begin{minipage}[b][1in][c]{1.5in}
	{\centering{\bf{Dr. Santanu Kumar Rath }}is a Professor in the Department of Computer Science and Engineering, NIT Rourkela since 1988. His research interests are in Software Engineering, }\\\\
\end{minipage}
  System Engineering, Bioinformatics, Natural Language Processing and Management. He is a Senior Member of the IEEE, USA and ACM, USA and Petri Net Society, Germany.\\\\
\end{document}